\documentclass{aa}  

\usepackage{graphicx}
\usepackage{txfonts,textcomp}
\usepackage[colorlinks=true, breaklinks=true, linkcolor=blue, urlcolor=cyan, citecolor=blue]{hyperref}

\usepackage{placeins}   
\usepackage{orcidlink}
\newcommand{\cms}{$\text{cm s}^{-1}$}
\newcommand{\kms}{$\text{km s}^{-1}$}

\begin{document} 
   \title{\texttt{SOAPv4}: A new step toward modeling stellar signatures in exoplanet research}
   \author{ E. Cristo\inst{\ref{inst1}}\orcidlink{0000-0001-5992-7589},
            J. P. Faria\inst{\ref{inst2}}\orcidlink{0000-0002-6728-244X},
            N. C. Santos\inst{\ref{inst1}, \ref{inst3}}\orcidlink{0000-0003-4422-2919}, 
            W. Dethier\inst{\ref{inst1}}\orcidlink{0009-0007-3622-0928},
            B. Akinsanmi \inst{\ref{inst2}},
            A. Barka\inst{\ref{inst1}, \ref{inst3}}\orcidlink{0009-0006-1182-7208},
            O. Demangeon \inst{\ref{inst1}}\orcidlink{0000-0001-7918-0355},
            J. P. Lucero \inst{\ref{inst1}, \ref{inst3}} \orcidlink{0000-0002-0080-4565},
            A. M. Silva\inst{\ref{inst1}, \ref{inst3}} \orcidlink{0000-0003-4920-738X}
            }

\authorrunning{E. Cristo et~al.}

   \institute{ Instituto de Astrofísica e Ciências do Espaço, Universidade do Porto, CAUP, Rua das Estrelas, 4150-762 Porto, Portugal\label{inst1}  
   \and
    Département d’astronomie de l’Universit\'e de Gen\`eve, Chemin Pegasi 51, 1290 Versoix, Switzerland \label{inst2}
    \and
   	Departamento de F\'{\i}sica e Astronomia, Faculdade de Ci\^encias, Universidade do Porto, Rua do Campo Alegre, 4169-007 Porto, Portugal\label{inst3}
             }

   \date{Submitted: 16th April, 2025 Accepted: 13th July, 2025}
 
  \abstract
   {In the era of high-resolution spectroscopy, methods for characterizing exoplanets and their atmospheres are advancing rapidly. As these techniques are refined and allow for the detection of even the most minute signals from the planet, however, the role of the host star becomes increasingly significant. The characterization of planetary systems relies not only on methods targeting the planet itself, but also on a detailed understanding of the host star and its activity at the spectral level.}
   {We present and describe a new version of the spot oscillation and planet code, \texttt{SOAPv4}. Our aim is to demonstrate its capabilities in modeling stellar activity in the context of RV measurements and its effects on transmission spectra. To do this, we employed solar observations alongside synthetic spectra and compared the resulting simulations.}
   {We used \texttt{SOAPv4} to simulate photospheric active regions and planetary transits for a Sun-like star hosting a hot Jupiter. By varying the input spectra, we investigated their impact on the resulting absorption spectra and compared the corresponding simulations. We then assessed how stellar activity deforms these absorption profiles. Finally, we explored the chromatic signatures of stellar activity across different wavelength ranges and discussed how such effects have been employed in the literature to confirm planet detections in radial-velocity measurements.}
   {We present the latest updates to \texttt{SOAP}, a tool developed to simulate active regions on the stellar disk while accounting for wavelength-dependent contrast. This functionality enables a detailed study of chromatic effects on radial-velocity measurements. In addition, \texttt{SOAPv4} models planet-occulted line distortions and quantifies the influence of active regions on absorption spectra. Our simulations indicate that granulation can introduce line distortions that mimic planetary absorption features, potentially leading to misinterpretations of atmospheric dynamics. Furthermore, comparisons with ESPRESSO observations suggest that models incorporating non-local thermodynamic equilibrium effects provide an improved match to the absorption spectra of HD 209458 b, although they do not fully reproduce all observed distortions.}
   {}


   \maketitle
%
\section{Introduction}
The connection between planets and their host stars is fundamental for characterizing planetary systems in detail. Stars are dynamic celestial bodies, and the signals we observe, both in terms of photometric and radial velocity (RV) studies, are equally complex. The main photospheric sources of these signals are stellar spots, faculae, and granulation for solar-like stars. Time series of active stars that are obtained with different techniques can be used to understand and infer several stellar properties, such as the velocity of the star \citep[e.g.,][]{2017A&A...603A...6N, 2024FrASS..1156379S} or its magnetic cycle \citep[e.g.,][]{2011arXiv1107.5325L,2012A&A...541A...9G, 2016A&A...595A..12S}. In planetary studies, it is a significant challenge to separate these signals. A key research focus therefore is understanding the effect of various stellar phenomena on the detection and characterization of exoplanets.\par
Stellar activity can either mimic or hinder signals of planetary origin in photometry and RV measurements for the characterization of the planet \citep[e.g.,][]{2010A&A...512A..38L, 10.1093/mnras/stu2730, andersen2015stellar} or its atmospheric detection \citep[e.g.,][]{2018ApJ...853..122R, 2018AJ....156..189C,2024A&A...685A.173C}.  To remove signals of stellar activity such as spots and faculae, and in particular, their effect on the modulation of RVs and photometry, the community often relies on the use of mathematical methods, such as Gaussian processes. These methods have yielded remarkable results, including the detection of L 98-59 b, a rocky warm planet with half the mass of Venus \citep{2021A&A...653A..41D}, and Proxima d, an Earth-mass planet orbiting the closest star to the Sun \citep{2022A&A...658A.115F}. Even though Gaussian processes are now a ubiquitous tool in such studies, their limited physical motivation and the wide range of possible kernel choices highlight the need for further investigation to quantify potential biases and the risk of spurious detections. An additional challenge lies in mitigating the effects of stellar granulation. In photometry, granulation has been recognized as a limiting factor in the detection of Earth-like planets that transit their host stars \citep[e.g.,][]{2020A&A...634A..75B,2020A&A...636A..70S}. In RV measurements, the issue is equally critical. For a solar-Earth analog system, the granulation-induced RV signal is expected to reach approximately 80~\cms \citep{2015A&A...583A.118M}\footnote{This amplitude represents a value that is unbinned in time and that decreases to $\sim$40\,cm\,s$^{-1}$ with a realistic cadence \citep{2015A&A...583A.118M,2020A&A...636A..70S}, at which point supergranulation contributes comparably \citep{2020A&A...642A.157M,2023A&A...669A..39A}.
}, whereas the planetary signal is anticipated to be only about 9~\cms. \par
The effect of stellar activity on the characterization of exoplanet atmospheres remains poorly explored. The majority of existing models is designed for low-resolution studies, in particular, spectrophotometric observations \citep[e.g.,][]{Macdonald2017,2018ApJ...853..122R, 2020A&A...635A.123B, 2023JOSS....8.4873M, 2025AJ....169...38Z}. While these models provide valuable insights into stellar activity \citep[e.g.,][]{2013A&A...549A..35O, 2014ApJ...791...55M, 2024A&A...692A..83P}, their applicability to high-resolution studies remains limited. Furthermore, systematic studies of the impact of stellar activity on transmission spectra are lacking, and only a few works addressed specific aspects such as the effect of unocculted spots in low resolution by \citet{2018ApJ...853..122R} and the influence on a limited set of high-resolution spectral lines by \citet{2018AJ....156..189C}. A comprehensive systematic framework is needed to assess the full spectral impact of stellar activity.\par
Quiet stars also pose problems for planetary characterization and have to be accurately modeled, in particular, during transits. As a planet moves across the stellar disk, it occults stellar regions with varying local stellar spectra. These variations in the observed local spectra over the surface of a star are known as the center-to-limb variations (CLVs). The main sources of these changes include the projected stellar rotation\footnote{This phenomenon is commonly, though inaccurately, referred to in the literature as the Rossiter-McLaughlin effect.}, the differing line profiles, and the flux variations (also known as limb-darkening). The latter two effects result from observations of stellar regions at different depths within the photosphere. It has been shown that CLVs are important to properly model the planet-occulted spectra during transits \citep{2020A&A...635A.206C,2021A&A...647A..26C,2023A&A...674A..86D}, with implications for the proper interpretation of planet transmission spectra.\par
Several studies have aimed to model these processes, and most adopted numerical approaches. Notable examples include previous \texttt{SOAP} versions \citep{2012A&A...545A.109B, 2013A&A...549A..35O, 2014ApJ...796..132D, 2018A&A...609A..21A, 2020MNRAS.493.5928S, 2024A&A...682A..28C}, which simulated the effects of stellar activity and planetary transits, the code called evaporating exoplanets (\texttt{EVE}) \citep[e.g.,][]{2013A&A...557A.124B, 2023A&A...674A..86D}, which is able to simulate the effect of atmospheric escape in exoplanets, and \texttt{StarSim} \citep{2016A&A...586A.131H, 2020A&A...641A..82R}, which models stellar activity and its impact on observed spectra. Additionally, the code called numerical empirical Sun-as-a-star integrator (\texttt{NESSI}) \citep{2024IAUS..365..389P} uses solar observations to compute solar-integrated spectra and study the impact of activity.\par
We present a new version of \texttt{SOAP}, hereafter \texttt{SOAPv4} \footnote{The code is publicly available on : \url{https://github.com/EduardoCristo/SOAP-Spot-Oscillation-And-Planet-code}}. This code builds on previous versions by adding the possibility to add real spectra for each stellar region. This allows us to simulate the spectroscopic and photometric time series of a planet that transits its host star and the signature of stellar activity.\par
This paper is structured as follows. Sect. 2 discusses the previous evolution of \texttt{SOAP} and details the new features that were introduced over time and in this version in more detail. In Sect. 3 we present applications of \texttt{SOAPv4} that show its potential for planetary transit modeling, including comparisons of absorption profiles with solar observations, analyses of the transit spectroscopy dataset of \object{HD~209458~b}, and its capability to model active regions at the spectral level. Finally, Sect. 4 summarizes the key aspects of this new version, highlights its limitations, and outlines prospects for future development.\par

\section{\texttt{SOAPv4}}\label{sectsoap}
\subsection{\texttt{SOAP} development through time}

The code \texttt{SOAP} \citep{2012A&A...545A.109B} was initially developed to simulate the impact of stellar activity (e.g., starspots) on RV time series.  The code simulates this by approximating the local stellar disk spectrum using a cross-correlation function\footnote{A CCF can be interpreted as a representation of the average line profile within a given spectral region.} (CCF), modeled as a Gaussian profile with user-defined parameters. The CCFs at different locations on the stellar surface are Doppler-shifted according to local velocities, assuming solid-body rotation, and flux-weighted following a limb-darkening law. Active regions are simulated by initially placing a circular region, defined in terms of the stellar radius, at the disk center. This region is then mapped to the specified latitude and longitude using spherical symmetry. The same Gaussian profile is applied to the CCFs of active regions. In terms of the flux contribution, the CCFs of active regions are weighted by their contrast: Values between 0 and 1 for dark spots, and values greater than 1 for bright faculae
\footnote{A note on terminology: In some previous works, the term "plages" was used instead of "faculae". While these terms are sometimes used interchangeably in the planetary community, in this code, we specifically simulate photospheric regions that are usually associated with faculae and not their chromospheric counterparts \citep[e.g.,][]{1982GAM....21.....P,2021isma.book.....B}.}.\par
Each successive version of SOAP was then built upon the previous framework. \texttt{SOAP-T} \citep{2013A&A...549A..35O} incorporated the effect of planet orbiting the star. In particular, the simulation accounts for the planet-occulted stellar regions during transits by subtracting the quiet-star (QS) pixels behind the planet at each position in the time series. This enhancement allows the code to simultaneously model the photometric light curve and the Rossiter-McLaughlin effect \citep{Holt1893,Rossiter1924,McLaughlin1924} in addition to the Keplerian motion. This expands its applications to understand the architecture of planetary systems better.\par
\texttt{SOAP2.0} \citep{2014ApJ...796..132D} was built upon the first version of the code (without planetary transits) to refine the physics of active regions. This version enables the use of solar observations to simulate stellar activity, specifically, by incorporating data from the Fourier Transform Spectrograph (FTS) at the Kitt Peak Observatory. It includes observations of the QS at the disk center \citep{1998assp.book.....W} and an umbral region of a sunspot \citep{2005asus.book.....W}, which allows a more realistic modeling of the stellar activity, at least in the solar case. These observations account for line-profile deformations induced by stellar activity, including the effects of convective blueshift \footnote{Granulation introduces distortions in line profiles that vary across the stellar disk. In observed spectra such as those from the IAG solar atlas, these effects are at least partially imprinted on the data, including line asymmetries and a component of the convective blueshift. While these spectra capture the average impact of granulation, however, they do not account for its temporal variability, that is, the changes in the RV that are induced by evolving granular patterns.} and its inhibition in active regions. Additionally, this version incorporates facular limb brightening and a realistic contrast ratio for spots and faculae by computing the flux contrast based on the Planck distribution at different temperatures between active regions and the quiet photosphere. \texttt{SOAP2.0} remains limited to using CCFs as in previous versions, however, and it is unable to generate realistic spectra.\par
The subsequent versions of the code were not publicly released, but played a crucial role in the development of \texttt{SOAPv4}. The advancements from \texttt{SOAP-T} and \texttt{SOAP2.0} were integrated by \citet{2016A&A...593A..25O} to study the impact of stellar activity on the measured misalignment of exoplanets. Later, \citet{2018A&A...609A..21A} enhanced the code to simulate the effects of a ringed planet and rotational deformation on the photometric transit light curve and the RV signal in the Rossiter-McLaughlin effect. More recently, \citet{2020MNRAS.493.5928S,2024A&A...682A..28C} incorporated the effects of differential rotation \citep[e.g.,][]{2016MNRAS.461..497B}, while \citet{2024A&A...682A..28C} added the convective blueshift signal that can be observed during planetary transits \citep{2011ApJ...733...30S}.\par
In addition to the main development line of the code, a fork of \texttt{SOAP2.0}, \texttt{SOAP-GPU} \citep{2023A&A...671A..11Z}, was created to simulate spectra. This version focuses on refining the stellar activity modeling by incorporating improvements such as direct mapping of active regions to account for complex shapes and a more advanced treatment of variations in the convective blueshift profile across the stellar disk, based on the measurements of \citet{2019A&A...624A..57L}. Furthermore, the code has been optimized for GPU execution, which significantly improved the computational efficiency compared to the original version. The requirement of a GPU architecture poses a usability challenge for most users, however. Furthermore, \texttt{SOAP-GPU} does not include the possibility of modeling a transiting planet.

\begin{figure}[t]
    \centering
    \includegraphics[width=1.0\linewidth]{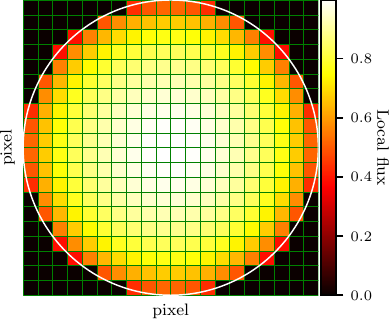}
    \caption{Depiction of the \texttt{SOAP} projection of a stellar disk onto a grid of $20 \times 20$ pixels for illustration purposes. The color map represents the local normalized flux in each position, which is determined by the limb-darkening. The white circumference outlines the original disk for reference}
    \label{fig:soap_grid}
\end{figure}
\subsection{The fundamental structure}\label{struct}
The \texttt{SOAP} code uses a numerical approach to simulate various aspects of stellar activity and planetary signals in astronomical measurements. In this new version, we have followed the main development trajectory by integrating all previous upgrades and extending its capabilities beyond photometry and CCFs to also model the time series of spectra.
The first step of the algorithm involves constructing the QS. This process involves projecting the stellar disk onto a square grid with side length $n$, resulting in a total of N ($n \times n$) grid points (see Fig. \ref{fig:soap_grid}). The next step selects the points within the stellar disk based on their position inside a unit disk, which is determined by calculating the distance between the centroid of each grid point and the center of the stellar disk. The position associated with each point is then used to compute the relevant physical quantities described below.\par
For each pixel in the grid, the flux contribution is determined by a limb-darkening law, and the corresponding coefficients are provided as inputs to the code. The total integrated flux at each time step is calculated by summing all grid pixels.\par
The signal of a transiting planet, after its relative position to the grid is computed, is obtained by recomputing the QS signal of the cells behind the planet and subtracting from the QS integrated signal.
For an active star, the flux contribution from active regions (ARs) is determined based on their position within the grid and its contrast relative to the QS. In the code, an AR is characterized by four parameters: latitude, longitude, radius ratio relative to the star, and temperature contrast.\par
Initially, an AR is placed at the center of the stellar disk, and its boundaries are defined by its radius. Subsequently, the ARs are translated into spherical coordinates according to the initial input parameters. In a time-series simulation, their coordinates are dynamically updated based on the observational phase, which is coupled to the stellar rotation.\par
\begin{figure}
    \centering
    \includegraphics[width=1.0\linewidth]{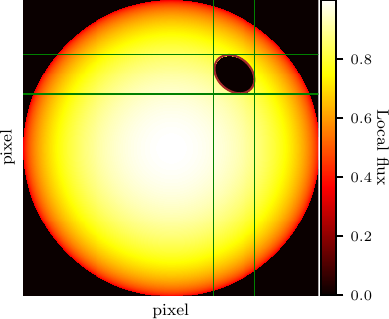}
    \caption{Position of a spot with a radius of $15\%$ of the stellar radius at a latitude and longitude of 45\textdegree and projected onto a $300\times300$ pixel grid. The vertical and horizontal green lines indicate the precomputed bounding box that marks the minimum and maximum Cartesian coordinates within the grid. The brown points represent the calculated positions that define the spot boundary in its final projected location.}
    \label{fig:spot_rep}
\end{figure}
At each time step, the code uses a precomputed bounding box around the AR position (illustrated in Fig. \ref{fig:spot_rep}). To determine the effect of the AR on the total integrated stellar signal, the code iterates over all grid points within the precomputed bounding box. Because the AR shape becomes distorted after projection onto the observed stellar disk, however, its precise geometry in the final coordinate system is unknown. To address this, the code applies an inverse rotation, for which it effectively shifts the AR back to the disk center, where its shape is well defined. This approach ensures that the QS signal can be accurately computed and subtracted from the original integrated stellar signal and that the specific contribution of the AR is isolated. Finally, the signal at the AR position is recalculated by multiplying the contrast between the star and the AR (calculated using Planck's law) with the limb-darkening effect \citep{2012A&A...545A.109B, 2014ApJ...796..132D}. For a transiting exoplanet, the code includes a condition to exclude the contribution of any AR located behind the planet. This optimization improves computational efficiency because the signal from these ARs would ultimately be removed.\par
\subsection{What is new in \texttt{SOAPv4}}
With \texttt{SOAPv4}, time-series simulations of CCFs and spectra are inherently linked. Since a CCF is derived as a weighted average of spectral lines based on a reference mask \citep[e.g.,][]{1996A&AS..119..373B,2002A&A...388..632P}, its computation is directly tied to the underlying spectra. Given this relation, we primarily focus on describing how the code simulates spectra, while details of CCF calculations can be found in \citet[][]{2012A&A...545A.109B} and \citet{2014ApJ...796..132D}.\par
A summary with the key differences between the versions is provided in table \ref{tab:soap_versions_structured}.
\subsubsection{The quiet-star spectrum}
The QS spectrum of a simulated star can be computed by integrating the local spectra of the star in different disk positions. These in turn depend on three key inputs: the local spectra, the local velocities, and the flux distribution across the stellar disk. The local spectrum is mainly determined by stellar properties such as effective temperature, metallicity, surface gravity, and other atmospheric parameters.\par
\begin{figure}
    \centering
    \includegraphics[width=\linewidth]{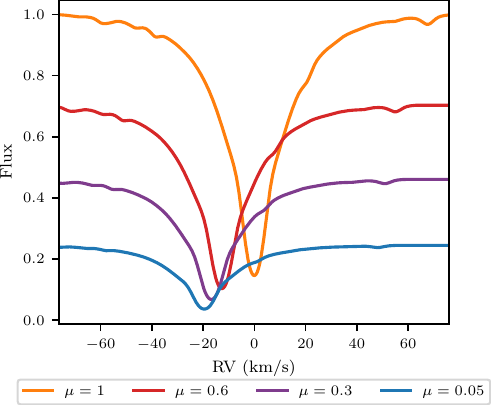}
    \caption{Local line profiles centered on the sodium $\text{D}_2$ line at different limb angles for a solar-like star that rotates as a solid body at 20 \kms. The representation illustrates velocity variations from the disk center to the eastern limb for a star rotating from west to east. As the pixels approach the limb, the flux decreases due to limb darkening, while the projected velocity component increases.}
    \label{fig:local_profiles_phoenix}
\end{figure}
\begin{figure}
    \centering
    \includegraphics[width=\linewidth]{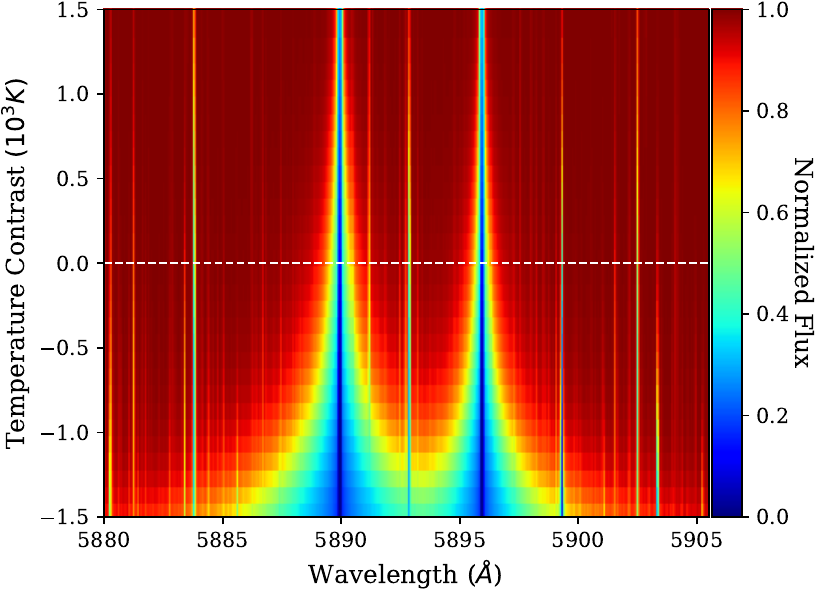}
    \caption{Variation in the sodium doublet line properties as a function of temperature contrast in regard to the solar case. Positive contrasts correspond to spectra of faculae, and negative contrasts represent spectra of spots. The quiet-star spectrum is indicated by the dashed white line at the zero level. All spectra used to construct the figure were taken from the PHOENIX library.}
    \label{fig:spectra_contrast}
\end{figure}
From a modeling perspective, each pixel in the grid must be assigned a spectrum that accurately reflects the local conditions of the star. To provide flexibility, the code allows users to specify this spectral information in various ways (Table \ref{tab:table_atmospheric} summarizes some of the options employed below). Users can input synthetic stellar spectra, such as those from the PHOENIX library \citep[\textsf{PHNX};][]{2013A&A...553A...6H}, or spatially resolved spectra across the stellar disk (e.g., generated with \texttt{Turbospectrum}; \citealt{2012ascl.soft05004P}), computed under either local thermodynamic equilibrium (LTE; model \textsf{TS-LTE}) or nonlocal thermodynamic equilibrium (NLTE; model \textsf{TS-NLTE}) conditions. Alternatively, observed spectra may be used, including solar spectra employed in the \texttt{SOAP2.0} CCF calculations for the quiet-Sun disk center (\textsf{FTS-QS}) or high-resolution spectra from the IAG atlas of the spatially resolved quiet Sun \citep[\textsf{IAG};][]{2023A&A...673A..19E}\footnote{The spectral library is  available at: \url{https://www.astro.physik.uni-goettingen.de/research/solar-lib/}}. Both the \textsf{IAG} and \textsf{TS} provide spectra as a function of disk position, capturing variations in the line profiles across the stellar surface. For the purposes of describing the code, we assumed that a single spectrum is applied to each grid element. This method can readily be extended to account for $\mu$-dependent spectra (where $\mu$ corresponds to the limb angle), however.\par
The input spectrum, which can optionally be normalized, is then convolved with the instrumental profile at a resolution specified by the user.\par
Each spectrum in the grid is Doppler shifted according to the local velocity of the star, incorporating effects such as differential rotation and convective blueshift, as described by \citet{2020MNRAS.493.5928S} and \citet{2024A&A...682A..28C}. Finally, the spectrum is interpolated back to the original wavelength grid, and its flux is rescaled according to the disk position using a wavelength-dependent limb-darkening prescription. This step is optional in series of $\mu$-dependent spectra where the intensities are known. A specific example of how local spectra can vary as a function of limb angle is shown in Fig. \ref{fig:local_profiles_phoenix}. The integrated quiet-star spectrum is then obtained by summing the contributions from all grid elements corresponding to the visible stellar disk.\par
\subsubsection{Including active regions}
For an active star, as described in Section \ref{struct}, the code identifies the region of the grid occupied by the active region (AR) or ARs at each time step and for each set of location angles. The activity component can consist of a single AR or a list of ARs, each with its own set of properties that evolve over time according to the stellar rotation. \par
The spectral contribution of the active region at each pixel is computed using a potentially different input spectrum (the choice is given to the user). One example is a synthetic spectrum, such as \textsf{PHNX}, which matches the temperature of the AR to capture spectral variations induced by temperature changes. In Fig.~\ref{fig:spectra_contrast}, we present the effect of temperature variations on the line profiles in the \ion{Na}{I} doublet region, adopting solar properties as the reference baseline (see Table~\ref{tab:hotjupiter}). The lower temperature of star spots leads to broader spectral lines than in the surrounding photosphere, whereas the higher temperature of faculae produces narrower lines. Consequently, cooler active regions predominantly affect the wings of the integrated spectral lines, while hotter regions exert a more significant influence on the line core.\par
When the spectral intensities are unknown, the flux contrast between the QS and the AR cannot be directly determined. In these cases, the code estimates the flux ratio by computing the ratio of the Planck distribution at a reference wavelength for the effective stellar temperature and that of the AR temperature, following a similar approach to that used at the CCF level.\par
For input spectra where intensities are known, \texttt{SOAP} operates with normalized fluxes. This is accomplished by normalizing the spectra to the maximum stellar flux within a specified wavelength interval\footnote{This normalization can be performed at any reference wavelength when the relative contrast is preserved and the wavelength is provided to the code.}, and scaling the AR spectrum to the same reference level at the corresponding wavelength. This wavelength is provided as an input parameter to the code to ensure that contrast is correctly maintained and that chromatic contrast is preserved in the simulations.\par
When a planetary transit is simulated, we follow the method described by \citet{2016A&A...593A..25O}, wherein the local signal of the QS behind the planet is computed and subsequently subtracted from the integrated QS signal. During the planetary transit, the AR signal is not calculated at the occulted positions to avoid unnecessary computational overhead.\par
Additionally, the RV signal, along with spectral indicators such as the full width at half maximum (FWHM) and bisector inverse slope (BIS), can be directly computed from the spectra using the CCF method. At the simulation level, the code allows users to input a cross-correlation template that specifies the wavelengths and weights, which can be a simulated spectrum, weighting function, or binary mask. The code then cross-correlates this template with the time series of the integrated spectra and produces a series of CCFs. The RVs are determined by fitting a Gaussian function to each CCF, while the FWHM and BIS indicators are computed directly from the CCFs.\par
To assess the performance of the code in simulating active regions, we compare the CCF results of \texttt{SOAPv4} with those of \texttt{SOAP2.0} in Appendix \ref{compare_vers}. In these simulations, we computed the variations in RV, BIS, and FWHM induced by two active regions with a 1$\%$ filling factor for a solar analog: a cool spot with a temperature 663 K below the photospheric value, and a facula whose temperature contrast follows the empirical law of \citet{2010A&A...512A..39M}, which is 34.1 K hotter than the photosphere at the disk center and increases to 209.5 K near the limb. The results are generally consistent, but a discrepancy is observed at the centimeter per second level. This difference has been verified to stem from modifications in the interpolation method and the selection of grid points. These adjustments are anticipated to yield more accurate results because they more effectively approximate the derivative of the line at the first point and, in the second instance, reduce the grid error to half a pixel. \par
\section{Applications}\label{app}
This section demonstrates the versatility and capabilities of the \texttt{SOAPv4} model through several key applications. We explore how this tool can be used to interpret transmission spectra, model stellar activity, and distinguish between stellar activity signals and other spectral features.
First, we examine the model performance in simulating the \object{HD~209458} system, with particular focus on the sodium (Na) doublet region where atmospheric detection has been debated in the literature (to be discussed in section \ref{hd209}). We then use spatially resolved \textsf{IAG} spectra to validate the model against solar benchmark data and assess the effect of different model parameters on the simulated signals.
The section also investigates the critical impact of stellar activity on transmission spectra by simulating spot-crossing events and analyzing their spectral signatures. We demonstrate that spots can create distortions in absorption profiles that might be misinterpreted as planetary atmospheric features. Finally, we show that \texttt{SOAPv4} can model chromatic RV variations caused by stellar activity. This makes it a valuable tool for distinguishing between stellar activity and genuine planetary signals.
Through these applications, we illustrate that \text{SOAPv4} serves as a comprehensive framework for investigating the complex interplay between stellar features and planetary signals in exoplanet characterization studies.\par
For the simulations presented in the following subsections, we adopted a grid size of 800, which is generally sufficient to accurately model transits of hot-Jupiter-sized planets and active regions with sizes of about $1\%$. Table \ref{tab:table_atmospheric} summarizes the key properties of the input spectra we used in these simulations.
\begin{figure}
    \centering
    \includegraphics[]{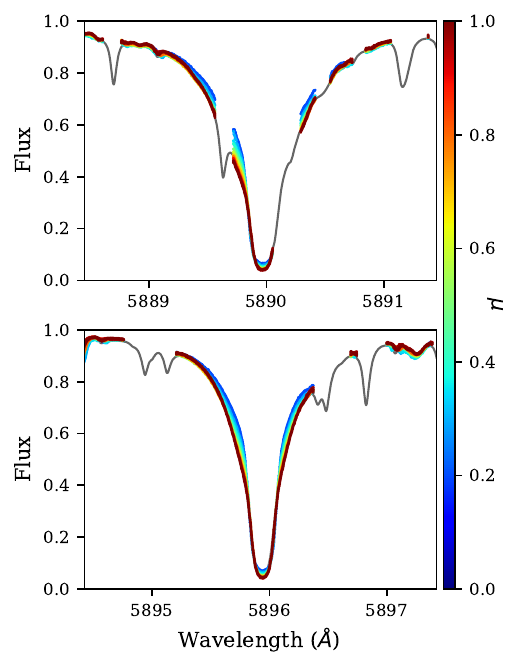}
    \caption{\textsf{IAG} \ion{Na}{I}~D$_2$ (top) and \ion{Na}{I}~D$_1$ (bottom) \ion{Na}{I} lines as a function of limb position. The wavelength interval was selected to represent a 3$\AA$ region surrounding each line. Contaminated regions that are affected by neighboring lines were excluded, as demonstrated by the comparison with the integrated star spectrum shown in black.}
    \label{fig:IAG_spectra_Na}
\end{figure}
\subsection{Solar simulations}\label{sect:solarsims}
Due to its proximity with Earth, the Sun provides a unique opportunity to gain detailed insights into its dynamics and to use it as a testbed for various models and extraction techniques. Furthermore, its observations can serve as a benchmark for the model validation because high-resolution observations of different regions of the solar disk allow direct comparisons. As new models emerge in the literature that depend on local phenomena in the stellar disks, it is crucial to first evaluate their performance against the solar case.\par
In this subsection, we employ spatially resolved solar observations from \textsf{IAG}, along with \textsf{PHNX} and $\mu$-dependent \textsf{TS-LTE} and \textsf{TS-NLTE}, to simulate the expected absorption signal caused by planet-occulted line distortions (POLDs) using \texttt{SOAPv4}. A similar study was reported by \citet{2023A&A...673A..71R} for the \object{HD~189733} system. These results, however, only offer an order of magnitude for this specific case, however, because the star is considerably cooler than the Sun ($4969\pm48$ K; \citealt{2021A&A...656A..53S}). We chose to model a scenario of a typical hot Jupiter orbiting a solar analog.\par
The \textsf{IAG} spectra result from observations using the Fourier Transform Spectrograph (FTS) at the Institut für Astrophysik und Geophysik (IAG) in Göttingen. The observation were conducted with a 32.5" fiber size, observing 14 $\mu$ positions that ranged from disk center to very close to the stellar limb. It is important to note, however, that because the physical size of the fiber used in observations is finite, each recorded profile is not strictly spatially localized. This effect becomes more pronounced toward the stellar limb, where the increased overlap between different $\mu$-angles results in greater deviations.\par
\subsubsection{Inactive solar analog}\label{QS_sun}
We investigated the POLDs associated with the \ion{Na}{I} D lines because they are among the most extensively studied alkali metal features in the optical regime \citep[e.g.,][]{2002ApJ...568..377C,2008ApJ...686..658S,2022MNRAS.514.5192L,articlesodium}. Their typically deep and broad profiles make them favorable for detection, and they have therefore been a prominent target in the literature.\par
We defined the physical parameters for a Jupiter-like planet orbiting at \(9\,R_\odot\), corresponding to an orbital period of approximately 3.13 days. The remaining stellar parameters were assumed to be solar and are listed alongside the planetary parameters in Table~\ref{tab:hotjupiter}. We modeled the star assuming rigid-body rotation throughout the simulations. For the solar spectra, we employed four distinct sources: \textsf{PHNX}, \textsf{IAG}, \textsf{TS-LTE}, and \textsf{TS-NLTE}.\par
For the simulation using \textsf{IAG} data (and, by extension, for those employing the other input spectra used for comparison), we focused on the \ion{Na}{I}~D$_1$ line. The \ion{Na}{I}~D$_2$ line is heavily contaminated by neighboring lines, which are masked in the atlas. Consequently, the line profiles are affected by this and possible leftover spurious signals, which may introduce bias in the conclusions that are drawn. For the  D$_1$ line, however, most of the line structure is preserved (see Fig.\ref{fig:IAG_spectra_Na}).\par
For the simulations, we used normalized spectra or prenormalized all input spectra. It is common practice to normalize spectra by the continuum to compute transmission spectra, in particular, because modern high-resolution spectrographs typically lack flux calibration. Caution is warranted, however, because the choice of local normalization method can significantly influence the resulting transmission spectrum.\footnote{In particular, the local continuum slope is lost, which primarily affects broader spectral regions, but its effect remains negligible for smaller wavelength bins.}. For consistency, we selected the same wavelength intervals around the sodium lines for normalization in this study.\par
For \textsf{PHNX} spectra, the normalization was performed automatically by \texttt{SOAPv4}. The code selected flux points that corresponded to the outermost $10\%$ of the chosen intervals and rejected any spectral lines within this subinterval. \textsf{IAG}, \textsf{TS-LTE}, and \textsf{TS-NLTE} provided prenormalized outputs.\par
\begin{figure}
    \centering
    \includegraphics[width=\linewidth]{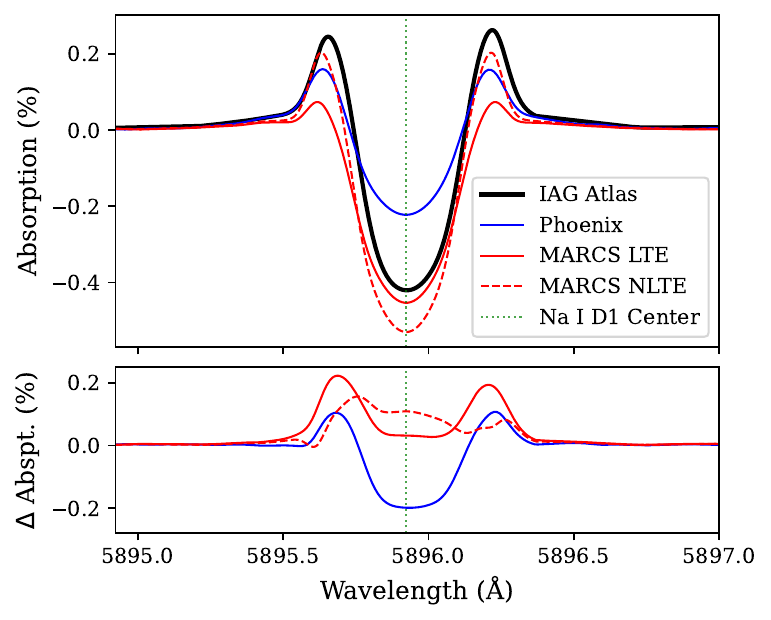}
    \caption{Comparison of the absorption spectra generated using \texttt{SOAPv4} for a solar analog hosting a hot Jupiter. The top panel presents a comparison of the absorption profiles obtained using input spectra from \textsf{IAG}, \textsf{TS}, and \textsf{PHNX} spectra that matched the stellar properties. The bottom panel illustrates the differences between the absorption spectrum produced with \textsf{IAG} and the spectra derived from the other models. These simulations represent the POLDs centered on the \ion{Na}{I}  D$_1$ line and span a wavelength range of 1$\AA$ on either side.}
    \label{fig:absorption_solar}
\end{figure}
In Fig. \ref{fig:absorption_solar} we show the resulting planetary absorption spectra\footnote{The absorption spectrum \( A(\lambda) \) is defined as the ratio of the in-transit integrated spectrum to the mean out-of-transit spectrum. It is related with the transmission through the relation $T(\lambda,t) = 1 - A(\lambda,t)$, where \( T(\lambda) \) represents the transmission spectrum. This formulation quantifies the fraction of stellar light that is absorbed by the planetary disk and the exoplanet atmosphere as a function of wavelength and provides insight into its composition and structure.}obtained from \texttt{SOAPv4} simulations using the different stellar spectra sources. The absorption simulation from \textsf{IAG} spectra exhibits the highest absorption at the wings. At the core, however, the emission is weaker than in the \textsf{TS-LTE} and \textsf{TS-NLTE} stellar atmosphere models. Neither the \textsf{TS} nor the \textsf{PHNX} simulations are able to reproduce the asymmetry that is observed with the \textsf{IAG} absorption spectrum. This suggests that the asymmetry is not driven by CLVs caused by the differences in line shape by varying atmospheric depths that are observed at different $\mu$-angles, nor is it caused by NLTE effects.\par
A key effect that is currently missing in \texttt{SOAPv4} and most models available in the literature, but s present in the IAG spectra due to their observational nature, is granulation \footnote{The granulation signal imprinted on the observed spectra}. Granulation distorts the stellar spectra locally due to the three-dimensional effect of observing granules from different perspectives, and it captures various components of the convective flows within the granular structure. This phenomenon has also been noted by \citet{2023A&A...673A..71R}.\par
Although CLVs do not explain the observed asymmetry, they remain an important factor in modulating the absorption signal. The constant \textsf{PHNX} model exhibits the largest deviation from \textsf{IAG} in terms of signal amplitude of the simulated profiles. The \textsf{TS} models provide a better approximation, but the LTE version retains residual absorption in the wings, whereas the NLTE version matches the \textsf{IAG} better. This indicates that NLTE effects cannot be neglected.\par
When we consider the \textsf{IAG}-simulated absorption profile as the reference model spectrum, it becomes evident that current modeling techniques face significant challenges. In particular, the \textsf{TS} LTE and NLTE models both overestimate the emission signal at the core. Because the absorption spectra are presented in the planetary rest frame, we might be inclined to attribute the observed signal to absorption originating from the exoplanet atmosphere. Furthermore, the asymmetry of the profile might be misinterpreted as evidence of atmospheric winds.
\begin{figure}
    \centering
    \includegraphics[width=\linewidth]{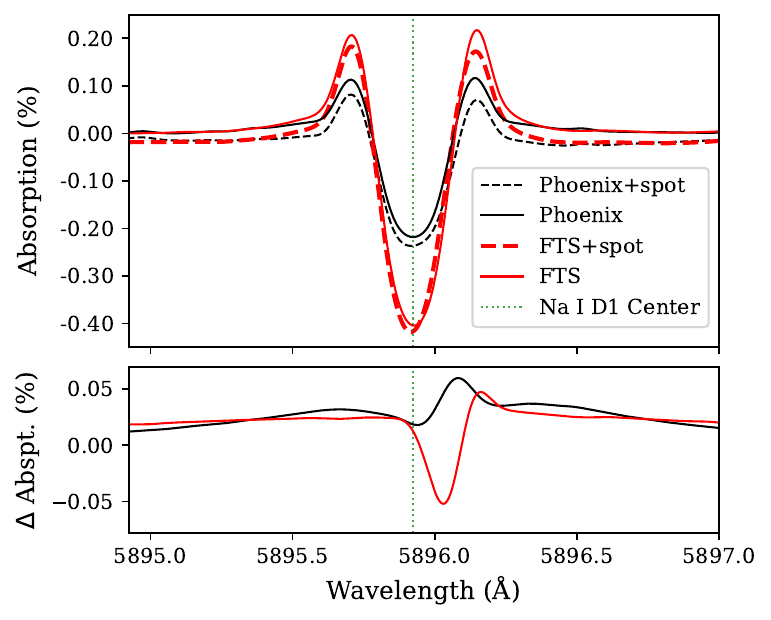}
    \caption{Top panel: Full absorption profiles for the \textsf{PHNX} (blue) and \textsf{FTS} (red) input spectra without active regions (solid lines), together with the corresponding profiles with a stellar spot (dashed lines).
Bottom panel: Differential signals obtained by subtracting the quiet-star profiles from the spot-crossing profiles. The color-coding matches the input spectra. The wavelength interval is the same as in Fig. \ref{fig:absorption_solar}.}
    \label{fig:spot_cross}
\end{figure}
\begin{figure}
    \centering
    \includegraphics[width=\linewidth]{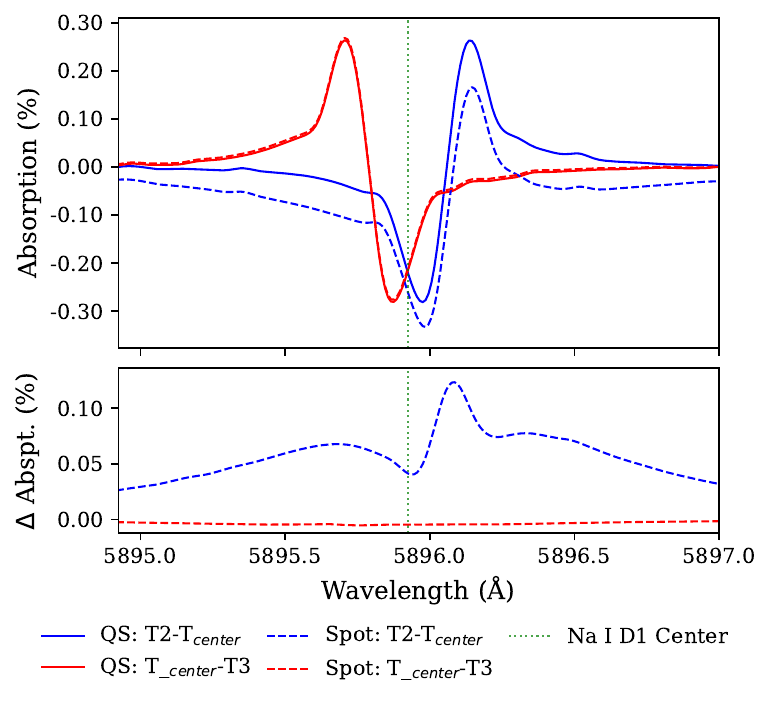}
    \caption{Comparison of partial mean absorption spectra for QS and spot-crossing events. In the top panel, the profiles for the two scenarios are presented. The leading hemisphere profiles (blueshifted) are represented in blue and the trailing hemisphere profiles (redshifted) are shown in red. The bottom panel shows the difference between the partial profiles for the quiet-star and spot-crossing events. The same wavelength interval as in Fig. \ref{fig:absorption_solar} is used.}
    \label{fig:spot_cross_sep}
\end{figure}
\subsubsection{Stellar activity}\label{AR_sun}
Stellar activity has recently become a key focus in the context of atmospheric characterization. As discussed in the introduction, most studies primarily investigated the impact of stellar activity on broadband transmission spectra or were limited in the number of features that were analyzed at higher resolution. \texttt{SOAPv4} serves as a valuable tool for exploring the effects of stellar activity on high-resolution transmission spectra by assessing how different active regions influence the extracted signals and determining how the resulting biases can be mitigated or accounted.\par
For these simulations, we considered a system as described in the previous section and set a spot-crossing event scenario. The simulations were based on \textsf{PHNX}spectra, where the properties of the QS spectrum are identical to those of the Sun, while the spot was modeled as being 663 K cooler than the stellar surface \citep{2011A&A...528L...9L}. Additionally, we tested simulating the absorption spectra with observed spectra from the FTS at the Kitt Peak Observatory, including spectra from the quiet Sun at disk center \citep{1998assp.book.....W} and an umbral sunspot region \citep{2005asus.book.....W} for comparison.\par
We simulated a starspot with a filling factor of $0.5\%$ at a latitude of -30\textdegree and a longitude of -35\textdegree, and we ensured that it was occulted during the transit. Furthermore, we introduced an impact parameter (b) of -0.6 to approximate the orbital inclinations of systems such as \object{HD~189733} or \object{HD~209458} (see the system architecture in Appendix \ref{app:hot_jupiter}, Fig. \ref{fig:hotjupiter}).\par
By default, \texttt{SOAPv4} computes and provides the transmission spectrum time series in the stellar rest frame by dividing the in-transit spectra by an out-of-transit spectrum. To shift it into the planetary rest frame, we first computed the planet velocities at the simulation phases and then applied a Doppler shift to remove these velocities, followed by an interpolation onto the original wavelength grid.\par
In Fig.~\ref{fig:spot_cross} we present a comparison between simulations with and without the spot contribution, using two input spectra: \textsf{PHNX} and \textsf{FTS}. Additionally, Fig.~\ref{fig:tomography} illustrates the time series of absorption spectra in the stellar rest frame for the \textsf{PHNX} spectrum simulation. Although the \textsf{PHNX} and \textsf{FTS} spectra are at the same temperature, the difference in amplitude between the simulations of the two models is significant. This is already visible at the input level of the spectra. The observed spectra have broader and deeper lines on average than the observations (Fig. \ref{fig:input_spec_compare}) \footnote{The resolution of the \textsf{FTS} observations is R>700,000, and that of the \textsf{PHNX} spectra is R=500,000.}\par
Regardless of the model, the subtraction of the scenario without a spot from the scenario with a spot reveals a distortion in the absorption spectrum that amounts to a significant percentage of the original absorption profile (see the lower panel of Fig. \ref{fig:spot_cross}). Locally, the continuum level of the spectral line appears to change and gradually approaches zero. This effect directly reflects the spectral profile of the star spot, which is broader than that of the quiet stellar surface. Over a broader wavelength range, however, broader than the width of the spectral spot lines, this apparent change in the continuum vanishes.\par
To the left of the line center, a redshifted feature appears in the difference between the model spectra. For the \textsf{PHNX} spectra, the quiet-star scenario exhibits a stronger absorption feature than the spot-crossing event. This outcome is expected to some extent because the local flux contrast distorts the line profile on the blueshifted side of the star.\par
In contrast, however, a similar structure appears in emission at approximately the same wavelengths in the FTS spectra. This discrepancy may arise from the relative behavior of the spectral lines: while the QS spectrum in \textsf{PHNX} is shallower than that of the spot, the opposite is observed in the FTS spectra.\par
This feature is well constrained within the absorption spectrum, and this may pose challenges for interpreting planetary atmospheric signatures in active stars. Its core is shifted from the expected velocity by less than 10 \kms, which is comparable to typical atmospheric dynamics. Repeated transit observations might help us to distinguish the origin of these deviations, however, provided that active regions evolve significantly between observations.\par
It is crucial to emphasize that we simulated a highly specific scenario. Variations in the location of active regions and differences in the orbital architecture of exoplanetary systems are expected to affect the resulting absorption profiles significantly. This underscores the importance of developing computational tools capable of assessing the contamination of planetary signals by stellar activity. In particular, the objective of the code \texttt{SOAPv4} is to provide a robust and flexible framework for a comprehensive analysis of this issue.\par
A single transit observation may still allow us to identify a spot-crossing event at the level of the absorption spectrum. Photometric and spectroscopic transit observations are often combined to enhance the detection capabilities. These complementary data are not always available, however, and it can be challenging to identify spot crossings based on RV measurements alone. Nevertheless, the residual asymmetry in Fig. \ref{fig:spot_cross} correlates with the spot position on the stellar disk.\par
Typically, constructing the absorption spectrum involves averaging the local absorption profiles along the transit chord, possibly excluding the ingress and egress regions, as discussed above. Under certain conditions, however, it may be feasible to separately average the profiles from T2-T$_\text{center}$ and from  T$_\text{center}$-T3. In a scenario that is perfectly parallel to the stellar equator, such as the we presented here, the resulting absorption spectra should be fully antisymmetric because the planet covers regions with the opposite velocities and spectral profiles. To test this, we divided the absorption profiles in these intervals and compared them with the QS scenario (Fig. \ref{fig:spot_cross_sep}).\par
The simulated spot was located on the leading stellar hemisphere, corresponding to the regions probed by the planet at T2-T$_\text{center}$ (blue lines in the figure). The effect of the spot crossing is clearly stronger than in Fig. \ref{fig:spot_cross}, primarily because of two factors. First, the averaging was performed exclusively in the blueshifted region of the star, which resulted in spectral lines with more similar positions. Second, the fraction of the spot signal relative to the overall absorption signal is larger. Compared to the QS scenario, a blueshifted signature is again observed in the difference between the two profiles. Furthermore, the local continuum exhibits additional absorption, which arises from the broader profile of the spectral lines of the spot. No significant features are observed in the trailing hemisphere. The local continuum does not reach the zero level because contamination from the spot remains present in the background from the construction of the master-out spectrum.\par
In the presence of stellar activity, the choice of the reference spectrum for comparison becomes particularly important. If the ratio of the transit duration and the stellar rotation speed is low (typically the case for slow rotators), there change in the velocity during observations is minimal to displace the spectrum of the active region. As a result, the contamination from active regions remains quasi-static and is effectively averaged out (but not removed, as we have shown) over the quiet star. When this ratio is sufficiently high, however dynamical effects may become apparent and superimpose themselves on the POLDs. They might even manifest themselves in out-of-transit observations.\par
This effect can be exacerbated by the method that is currently used to construct transmission spectra. Typically, out-of-transit integrated stellar spectra are aligned and corrected for systematic velocity trends by applying RV corrections that are derived from a fit to the observed velocities. Active regions not only induce genuine Doppler shifts, however, but also introduce distortions in the spectral lines. These distortions can be misinterpreted as velocity shifts, in particular, when methods such as the cross-correlation function (CCF) are used, and this might lead to a misalignment of the master-out spectrum. This misalignment can introduce spurious effects in the in-transit comparison and might distort the inferred transmission spectrum.\par
\subsection{Transmission spectrum of HD 209458}\label{hd209}
\begin{table}[ht]
\centering
\caption{\object{HD 209458}. Stellar and planetary properties.}
\label{tab:stellar_planetary_properties}
\begin{tabular}{p{5.5cm}c}

\textbf{Properties} & \textbf{Values} \\
\hline
\hline
\\
\multicolumn{2}{c}{\textbf{Stellar}} \\

Radius (R$_{\odot}$) & $1.158 \pm 0.0338$ \\
Mass (M$_{\odot}$) & $1.118 \pm 0.005$ \\
$T_{\text{eff}}$ (K) & $6126 \pm 18$ \\
Metallicity [Fe/H] & $0.04 \pm 0.01$ \\
$\log g$ (cm s$^{-2}$) & $4.368 \pm 0.007$ \\
$v_{\text{eq}} \sin i$ (km s$^{-1}$) & $4.228 \pm 0.007$ \\
\\
\multicolumn{2}{c}{\textbf{Planetary}} \\

Radius (R$_*$) & $0.121 \pm 0.004$ \\
Mass (M$_{\oplus}$) & $216.7^{+4.8}_{-4.4}$ \\
Semi-major axis (R$_*$) & $8.87\pm 0.05$ \\
Period (days) & $3.52474859(38)$ \\
Inclination (deg) & $86.71 \pm 0.05$ \\
$\lambda$ (deg) & $1.58 \pm 0.08$ \\
\hline
\end{tabular}
\tablefoot{The stellar parameters were taken from \citet{2021A&A...656A..53S}, with the exception of $v_{\text{eq}} \sin i$. This latter and the values for the planetary parameters were taken from \citet{2021A&A...647A..26C}. The planetary radius and mass were converted into units that are compatible with the input requirements for \texttt{SOAPv4}.}
\end{table}
\object{HD~209458~A} is a bright solar-type star with a visual magnitude of 7.63 \citep{2000A&A...355L..27H} and is classified as spectral type F9V \citep{2001AJ....121.2148G}. It hosts the exoplanet \object{HD~209458~b}, whose size is similar to that of Jupiter, but has only two-thirds of its mass. The favorable planet-to-star radius ratio, along with the extended nature of its atmosphere, makes \object{HD~209458~b} one of the most important targets for atmospheric characterization studies.\par
\begin{figure*}[t]
    \centering
    \includegraphics[width=\linewidth]{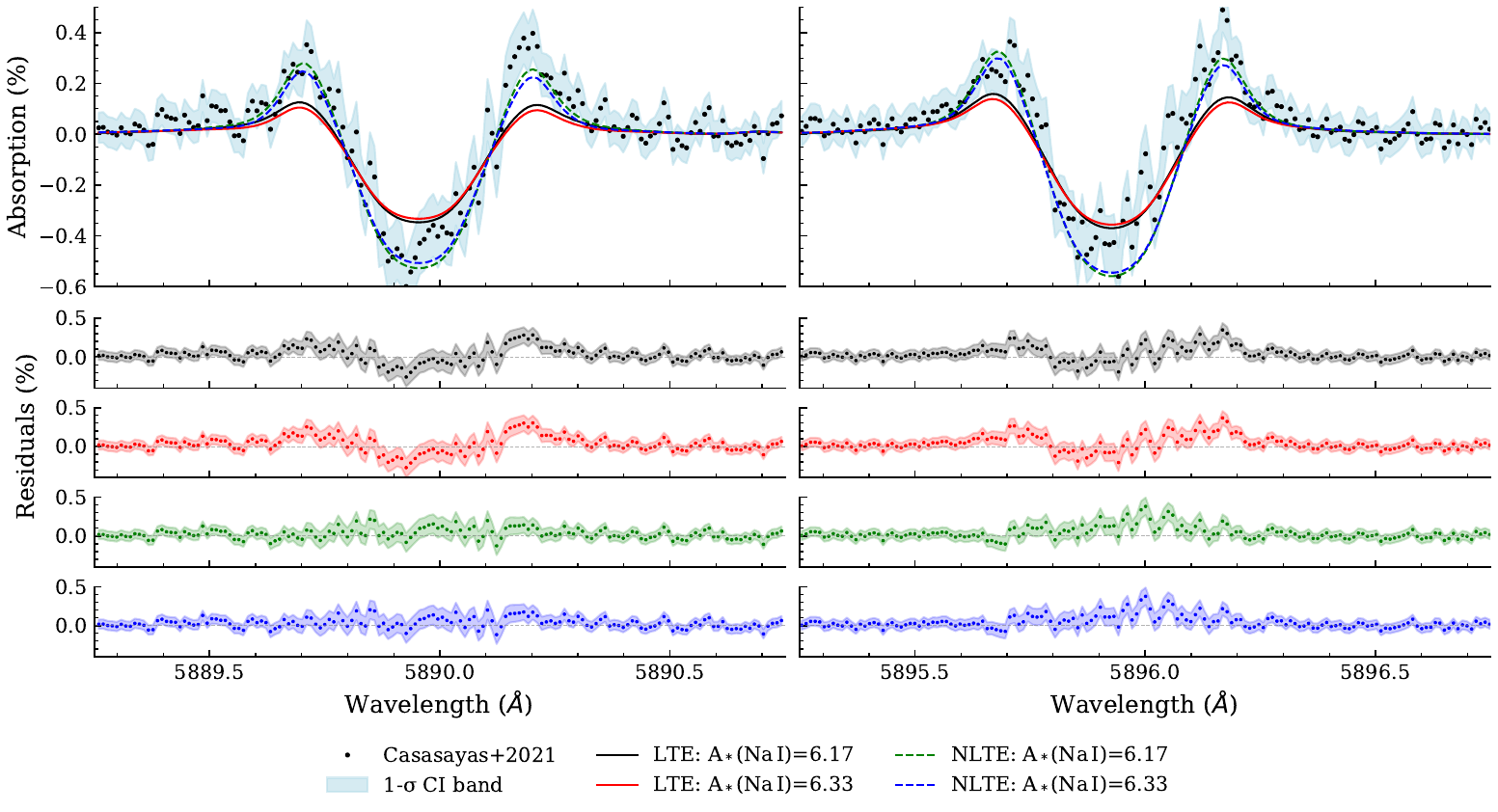}
    \caption{Absorption profiles of the Na\,I D lines in HD~209458 compared against the observed absorption profile derived from ESPRESSO data, as presented in \citet{2021A&A...647A..26C}. In the top row, the absorption points are shown in black, and their respective uncertainties are represented by a confidence interval (CI in the plot) band at the 1$\sigma$ level. The absorption features computed using \texttt{SOAPv4} are overplotted. They were based on input MARCS models with two Na\,I abundances (solar-like) reported in the literature. The subsequent rows display the residuals between the models and observations. The colors of the data points and CI bands correspond to the respective model color.}
    \label{fig:transmission_D2}
\end{figure*}
This planet was originally proposed to host the first detected atmosphere, inferred from an excess absorption in the sodium doublet region reported by \citet{2002ApJ...568..377C}. Since then, the authenticity of this detection has been a subject of ongoing debate. In low-resolution studies, \citet{2008ApJ...686..658S,2008A&A...487..357S, 2020A&A...644A..51S} confirmed this detection, whereas \citet{2020A&A...635A.206C, 2021A&A...647A..26C} showed no evidence of an atmospheric signature in high-resolution observations. More recently, \citet{2024A&A...683A..63C} showed based on simulations that the absorption feature that was observed in low-resolution by \citet{2002ApJ...568..377C} could not be explained by the effect of stellar rotation on the transmission spectrum. A plausible resolution to the paradox is that sodium absorption in the atmosphere of \object{HD~209458~b}mainly arises from the pressure-broadened wings of the \ion{Na}{I} D doublet, which formed in deeper, high-pressure layers. These broad wings can span several nanometers and appear as shallow but wide features in low-resolution spectra, which makes them detectable over large bandpasses. In contrast, high-resolution spectra isolate the narrow-line cores, where absorption may be weak or absent because the Na abundance at high altitudes is low or because it is obscured by clouds or hazes, resulting in little to no signal at these wavelengths. In ground-based high-resolution observations, this absorption thus disappears as a result of the continuum normalization on the data.\par

We used \texttt{SOAPv4} to simulate the sodium absorption feature that is expected from the deformation of the stellar lines by the planet during the transit.  An accurate modeling of the stellar surface behind the planet is essential because the range of planetary velocities that are projected along the line of sight during the transit causes any potential atmospheric absorption signal to fall within the velocity range of the POLDs. Because the host star is in the hotter range of solar-type stars, it has been shown \citep{2021A&A...647A..26C,2023A&A...674A..86D} that the the POLDs in high-resolution absorption spectra can only be accurately reproduced when NLTE effects in the spectral lines synthesis are accounted for. Therefore, we used synthetic spectra generated under NLTE conditions for the simulations to investigate the effect on the transmission spectrum.\par
To derive the transmission spectra from our simulations, we adopted the parameters listed in Table \ref{tab:stellar_planetary_properties}. One of the key parameters that significantly affects the amplitude of the POLDs is the stellar rotation speed, $v_{\text{eq}} \sin i$ \citep{2023A&A...674A..86D}. This parameter can be determined through various techniques, including stellar line broadening \citep[e.g.,][]{gray_2005}, the analysis of stellar variability in photometric observations \citep[e.g.,][]{2013A&A...560A...4R, 2014ApJS..211...24M, 2024FrASS..1156379S}, and the Rossiter-McLaughlin effect \citep[e.g.,][]{2011ApJ...742...69H, 2017MNRAS.464..810B}.\par
Additionally, the rotation of the stellar surface can be inferred by analyzing the missing spectral information of a star during a planetary transit \citep{2017MNRAS.464..810B}. This technique has become increasingly common and is now frequently employed alongside transmission spectroscopy to refine the stellar parameters, such as the stellar rotation speed or differential rotation \citep[e.g.,][]{2020A&A...641A.120H, 2021A&A...647A..26C, 2023A&A...672A.134S}. One of the key advantages of \texttt{SOAPv4} is its ability to self-consistently model the Doppler-shadow signal \citep[e.g.,][]{2016A&A...588A.127C,Dravins_2017,Dravins_2018}, which enables us to recover the transmission spectrum based on this information.\par
For this analysis, and to take into account the change in line shape as a function of the limb angle, we used \texttt{Turbospectrum} to produce synthetic spectra for a specific array of $\mu$ angles to be provided to the code. As parameters, we used the stellar physical properties in Table \ref{tab:stellar_planetary_properties}. Similarly to \citet{2021A&A...647A..26C}, we tested models with both LTE and NLTE approximations, using a solar \ion{Na}{I}abundance, A$_*$(Na$\,$I). Two main values are used in literature: $6.33\pm0.03$ \citep{1998SSRv...85..161G} and $6.17\pm0.04$ \citep{2007SSRv..130..105G}. In Fig. \ref{fig:shadow} we show the map of the local spectra in a wavelength interval between 5882 $\AA$ and  5902 $\AA$, which encloses the \ion{Na}{i} D$_1$ and D$_2$ lines.\par
We note that during transit, the profiles behind the planet change significantly due to CLVs. In the literature and for an observational analysis, however, it is common to average the transmission spectrum over the entire transit (T1–T4) or restrict it to the fully in-transit phases (T2–T3) to increase the signal-to-noise ratio. A restriction of the phase range to T2–T3 ensures that the planet remains entirely within the stellar disk, which avoids positions in which the $\mu$ angles are very close to the stellar limb. This is often challenging to model. In our simulation, we fixed the profiles to $\mu$ = 0.2 when the angle fell below this threshold.\par
In Fig. \ref{fig:transmission_D2} we show the absorption spectrum as derived from ESPRESSO observations by \citet{2021A&A...647A..26C}  for the sodium  D$_1$ and  D$_2$ lines, compared with the models obtained using \texttt{SOAPv4}. The models generated with NLTE spectra agree better with the absorption profile, in particular, to explain the shape of the wings and the amplitude of the feature.\par
For the absorption feature associated with the  D$_2$ line, the model reproduces the left wing more accurately than the right wing, where an apparent excess absorption is observed relative to the model. A similar trend is seen for the  D$_1$ feature, with a pronounced excess absorption in the left wing that the models fail to capture. Additionally, near the line cores, an absorption feature extending over approximately 0.1$\AA$ is observed in both the  D$_1$ and  D$_2$ lines. A possible explanation for this excess absorption close to the absorption core was given by \citet{2023A&A...674A..86D}, who modeled the effect of a planetary atmosphere containing \ion{Na}{I}in its composition, which explained the additional absorption that is seen. In addition, the authors were able to explain the asymmetry that is observed by modeling an atmosphere with a thermosphere with a night-to-day side wind of around 3 \kms.\par
For this showcase scenario, the model computed with \texttt{SOAPv4} agrees well with the models used by \citet{2020A&A...635A.206C,2021A&A...647A..26C, 2023A&A...674A..86D} and can explain the observations to a similar level of accuracy.

\subsection{Application to RVs}\par
Modeling stellar activity is crucial for planet detection and characterization through the RV method \citep{2024arXiv240605447R} because it can obscure the signals of low-mass planets. Previous versions of \texttt{SOAP} addressed this by simulating the CCF at the pixel level, where the contrast between the quiet star and active regions was determined using a reference wavelength at which the Planck distribution ratio of the stellar and AR temperatures was computed.  Although the same approach can be applied to small segments of spectra, it is generally more practical to compute the CCF over the full bandpass of an instrument.\par
In \texttt{SOAPv4}, it is now possible to simulate large portions of the spectrum while maintaining the contrast ratio of the star and ARs. This is achieved by normalizing the spectra to the maximum stellar flux and using the corresponding wavelength to scale the AR spectrum to the same reference value. The code then internally recalculates the appropriate flux level when it determines the local AR contribution and takes the temperature into account.\par
To illustrate the variation in the spot RV amplitude with wavelength, we simulated a time series of integrated spectra for a solar-type star as it would be observed with ESPRESSO. We considered a spot at the disk center with a 1$\%$ filling factor and an effective temperature 663 K below the photosphere. In this case, we applied an average limb-darkening law derived using LDtK \citep{2015MNRAS.453.3821P} with the spectral library of \citet{Husser2013}, integrated over the ESPRESSO bandpass, which spans 3800 $\AA$ and  7880 $\AA$.\par
\begin{figure}
    \centering
    \includegraphics[width=\linewidth]{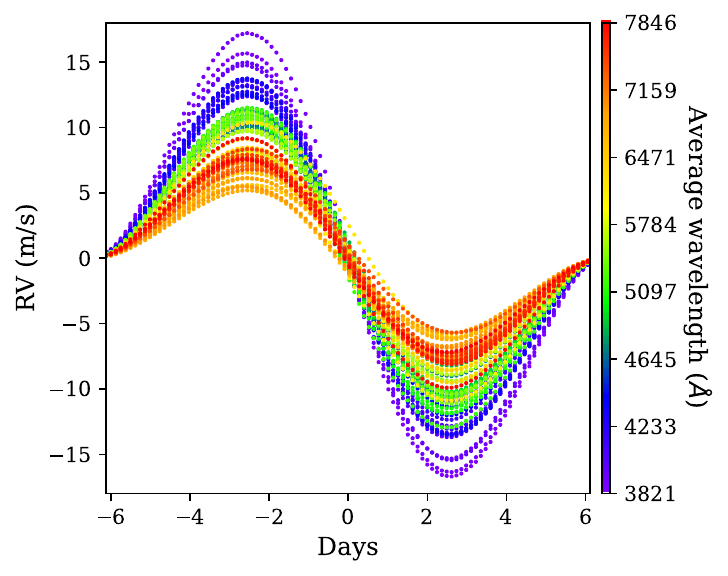}
    \caption{RV curves computed using \texttt{SOAPv4} for an equatorial spot with a $1\%$ filling factor, which is 663 K cooler than the photosphere. The color bar represents the average wavelength of the spectral bins, from which the radial velocities were derived using the CCF technique with the QS solar \textsf{PHNX} model.}
    \label{fig:chromatic_rvs}
\end{figure}
In Fig.\ref{fig:chromatic_rvs} we present the RV time series obtained by cross-correlating the solar \textsf{PHNX} model with several chromatic spectral bins of the active integrated star. These bins were constructed by dividing the initial interval into 70 intervals, each containing an equal number of points (with the exception of the last bin). A clear trend in amplitude is evident, with higher RV amplitudes observed at shorter wavelengths in the visible spectrum, and progressively lower amplitudes for longer wavelengths.\par
This trend in amplitude is known, and it is mainly a consequence of the energy distribution contrast between two sources at different temperature \citep[e.g.,][]{2008A&A...489L...9H, 2010ApJ...710..432R}. It was used as a tool to distinguish stellar activity from true planetary signals \citep[e.g.,][]{2008A&A...489L...9H,2020A&A...639A..77S,2022A&A...658A.115F}.
\section{Discussion and prospects}\label{disc}
We have presented the most recent updates to the code \texttt{SOAP} and its applications at the spectral level. The current version, \texttt{SOAPv4}, is capable of simulating the effects of ARs at any position on the stellar disk, assuming a circular shape for the ARs when they are at the disk center. Additionally, \texttt{SOAPv4} preserves the wavelength-dependent contrast, that is, the energy distribution difference between the star and the AR. This feature allows us to evaluate chromatic effects on RVs.\par
With respect to planetary transits, \texttt{SOAPv4} proved to be a valuable tool for modeling the POLDs under a variety of input spectra, whether synthetic or observed, and for assessing the impact of active regions on absorption spectra. As an illustrative example, we showed that for a solar-analog star hosting a representative hot Jupiter, absorption spectra modeled with solar observations cannot be fully reproduced by synthetic spectra that share the properties of the Sun. We suggest that the asymmetry that is observed may in part be due to the granulation effect on the solar surface that distorts the spectral lines locally. Moreover, the distortion observed in the comparison may mimic an asymmetric planetary absorption feature. These distortions might not only lead to false detections, but might also generate spurious signatures of atmospheric dynamics. We note, however, that the difference between these simulations of absorption spectra is a function of the system properties, and the amplitude and distortion of the signal are expected to change accordingly.\par
For the scenario involving an active region, we used the same illustrative example as before, with an orbital configuration that resulted in the hot Jupiter occulting the spot during the transit. In this case, we compared the effect on the POLDs of observed solar spectra from the disk center and the umbral region of a spot (\textsf{FTS-spot}), alongside \textsf{PHNX} spectra with matching stellar properties. Similarly as for the QS case, the \textsf{PHNX} spectra tended to underestimate the amplitude of the absorption spectra. In the relative comparison of absorption spectra with and without a spot-crossing, both scenarios produced a blueshifted feature, but the \textsf{PHNX} spectra showed relative emission, while the FTS spectra displayed relative absorption. These differences reflect not only the relative line intensities between QS and active region spectra, but also intrinsic differences in the initial line shapes. We further demonstrated that for an aligned system, a comparison of the average absorption spectra from the post-egress phase to mid-transit and from mid-transit to pre-egress phases can accentuate the distortion that is observed in the average transmission spectrum. This might serve as a tool for confirming spot-crossing events.\par
To demonstrate the capabilities of the upgraded \texttt{SOAPv4}, we compared the simulated absorption spectra for the \object{HD~209458} system with the absorption spectra derived from transit observations with ESPRESSO as reported by \citet{2021A&A...647A..26C}. The absorption spectra were better reproduced by synthetic models using NLTE conditions because LTE models tend to underestimate the amplitude of the absorption feature. These conclusions agree with those presented by \citet{2021A&A...647A..26C, 2023A&A...674A..86D}. In comparison to these studies, we observed a similar degree of agreement between the observations and the models. The NLTE models did not fully account for the absorption profile, however. As noted by \citet{2023A&A...674A..86D}, additional absorption is observed near the center of the profile, which may be attributed to planetary atmospheric absorption that is redshifted as a result of the dynamic processes in the exoplanet atmosphere. Nevertheless, we demonstrated in our illustrative example that the differences between models during a spot occultation can reproduce similar absorption features. Multiple observations of the same target during transit may help us to distinguish true absorption features from stellar activity contamination. As seen in the solar simulations presented in Sect.~\ref{sect:solarsims}, the left and right wings of the absorption profile closely resemble those obtained in our illustrative example using \textsf{IAG} spectra. This similarity suggests that granulation might contribute to the deformation of the absorption profile and should therefore be taken into account when modeling for an accurate interpretation of potential atmospheric signals.\par
In the case of the solar-analog star, we demonstrated that more accurate local synthetic models are required to explain the spectral observations behind the planetary transit. For stars with a convective photospheric layer, granulation effects might introduce significant line distortions. Therefore, integrating synthetic spectra from codes such as \texttt{CO5BOLD} \citep{2007AN....328..323S,2010ascl.soft11014F}, \texttt{MURaM} \citep{2024A&A...681A..81W} or \texttt{Stagger} \citep{2024ApJ...970...24S}, which simulate the effects of convection, might offer a solution to better account for these distortions.\par
In modeling active regions, the current code is limited in terms of the spectra that are available for use. For the simulations presented in Sect. \ref{AR_sun}, we assumed when we used synthetic spectra that the spot spectrum corresponds to the spectrum of a star with the same temperature as an average solar spot. This is merely an approximation, however, that might even contribute to the discrepancies observed between the model and the observations in Fig. \ref{fig:input_spec_compare}. Although ARs might be less prone to distortions due to partial inhibition of convection \citep{2017A&A...607A...6M}, they are associated with regions with strong magnetic fields that can create an observable Zeeman splitting of the spectral lines. Furthermore, the internal structure of spots (comprising the cooler umbra and the warmer penumbra) exists under distinct thermodynamic conditions that are not adequately captured by a single averaged spectrum. In addition, the position, physical properties, and temporal evolution of active region structures on stellar surfaces remain largely unconstrained.\par
Distinguishing the signals that arise from the POLDs and stellar activity alone presents a significant challenge because any absorption signal that originate from the exoplanet atmosphere may introduce bias in the interpretation. It is not a viable solution to model the atmospheric signal alone because it would be subject to the same biases. To address this, we plan to enhance the current version of the code we presented in this paper. Future improvements will involve simulating the simultaneous effects that originate from both the star and the planet by incorporating a three-dimensional atmospheric grid around the planet.\par
Beyond improvements in modeling, more robust local solar observations are essential. High-resolution measurements of the different structures within ARs are needed to better understand the temporal evolution of associations between spots and faculae. For local spectral profiles, it is also crucial to obtain spectra in smaller regions. For reference, the IAG atlas of the resolved Sun currently uses a spatial resolution of 32.5". This is particularly critical near the stellar limb, where the line properties change more rapidly and are significantly affected by averaging due to the fiber size. Future instruments, such as the Paranal Solar ESPRESSO Telescope (PoET; \citealt{2024SPIE13096E..74L,2025Msngr.194...21S}), will make significant contributions to this effort. Integrated with the ESPRESSO spectrograph, PoET will enable both disk-integrated and disk-resolved observations of the Sun at an ultrahigh spectral resolution (R $\sim$ 200,000). Moreover, its smallest fiber, with a size of 1", is comparable to the scale of a solar granule, which will allow an unprecedented spatial and spectral detail in solar observations.\par
\begin{acknowledgements}
This work was funded by the European Union (ERC, FIERCE, 101052347). Views and opinions expressed are however those of the author(s) only and do not necessarily reflect those of the European Union or the European Research Council. Neither the European Union nor the granting authority can be held responsible for them. The authors acknowledge X. Dumusque for his invaluable advice and discussion. BA acknowledges the financial support of the Swiss National Science Foundation under grant number PCEFP2\_194576. ODSD acknowledges support from e-CHEOPS: PEA No 4000142255. This work was supported by FCT - Fundação para a Ciência e a Tecnologia through national funds by these grants: UIDB/04434/2020 DOI: 10.54499/UIDB/04434/2020, UIDP/04434/2020 DOI: 10.54499/UIDP/04434/2020.
\end{acknowledgements}

\bibliographystyle{aa}
\bibliography{bib}

\begin{appendix} 
\onecolumn
\section{Summary of the \texttt{SOAP} versions}\label{vers_sum}

\begin{table}[!htbp]
\centering
\caption{Comparison of major SOAP versions by core capabilities and modeling details.}
\begin{tabular}{p{0.18\linewidth} p{0.23\linewidth} p{0.20\linewidth} p{0.04\linewidth} p{0.25\linewidth}}
\hline
\centering{\textbf{Version}} & \textbf{Input} & \textbf{AR Modeling} & \textbf{Planet} & \textbf{Star} \\
\hline
\centering{\texttt{SOAP} \\ ~\citep{2012A&A...545A.109B}}
& CCF (model)
& ARs' RV impact modeled by local CCFs with brightness contrast relative to the photosphere, Doppler-shifted by stellar rotation. The same contrast applies to flux variation.
& No
& Local flux weighted by linear limb-darkening law, adjusted for stellar properties. \\ \\

\centering{\texttt{SOAP-T} \\ ~\citep{2013A&A...549A..35O}}
& CCF (model)
& Inherited from \texttt{SOAP}
& Yes
& Includes planetary transits and occultations of ARs. \\ \\

\centering{\texttt{SOAP 2.0} \\ ~\citep{2014ApJ...796..132D}}
& CCF (Observed + model)
& Includes inhibition of convective blueshift in ARs (\textsf{FTS}), limb brightening for faculae, and realistic spot/plage contrast ratios (Planck distribution).
& No
& Flux weighted by quadratic limb-darkening law; observational convective blueshift suppression. \\ \\

\centering{\texttt{SOAP 3.0} \\ ~\citep{2018A&A...609A..21A,2020MNRAS.493.5928S}}
& CCF (Observed + model)
& Inherits \texttt{SOAP 2.0} AR modeling.
& Yes
& Simulates transits of ringed planets and their occultation of ARs; stellar surface with differential rotation. \\ \\

\centering{\texttt{SOAP-GPU} \\ ~\citep{2023A&A...671A..11Z}}
& Observed spectra (solar)\newline
Synthetic spectra (\textsf{PHNX})
& Quiet photosphere, spots, and faculae modeled using \textsf{PHNX} synthetic spectra at each $\mu$ angle. Direct mapping of the ARs.
& No
& GPU-based full-disk spectral synthesis; injects disk-position-dependent bisector profiles. \\ \\

\centering{\texttt{SOAPv4} \\ (\citealt{2024A&A...682A..28C}; This work)}
& CCF (Observed + model)\newline
Observed spectra (\textsf{IAG}, \textsf{FTS}, generic)\newline
Synthetic spectra (\textsf{PHNX}, \textsf{TS}, generic)
& Inherits \texttt{SOAP 2.0} AR modeling for CCF; adds spectral modeling.
& Yes
& Simulates spectral time-series including transmission spectra (POLDS), Doppler shadow profiles line-by-line or full spectrum, and activity impact on line profiles. \\
\hline
\end{tabular}
\label{tab:soap_versions_structured}
\end{table}
\section{Summary of the input spectra}\label{app:input_spectra}

\begin{table}[h!]
\centering
\caption{Summary of the input spectra used by \texttt{SOAPv4}.}
\begin{tabular}{llll}
\hline
\textbf{Label}     & \textbf{Source} & \textbf{Type}   & \textbf{Notes} \\
\hline
\textsf{FTS-QS}    & Observed                   & Disk Center         & QS\\
\textsf{FTS-spot}  & Observed                   & Disk Center         & Spot umbra\\
\textsf{IAG}       & Observed                   & $\mu$-array         & Quiet Sun only \\
\textsf{PHNX}      & PHOENIX                    & Disk Center         & Synthetic spectrum \\
\textsf{TS-LTE}    & MARCS                      & $\mu$-array         & LTE approximation \\
\textsf{TS-NLTE}   & MARCS                      & $\mu$-array         & NLTE treatment \\
\hline
\end{tabular}
\label{tab:table_atmospheric}
\end{table}
\section{Hot Jupiter properties and architecture}\label{app:hot_jupiter}

\begin{figure}
    \centering
    \includegraphics[width=\linewidth]{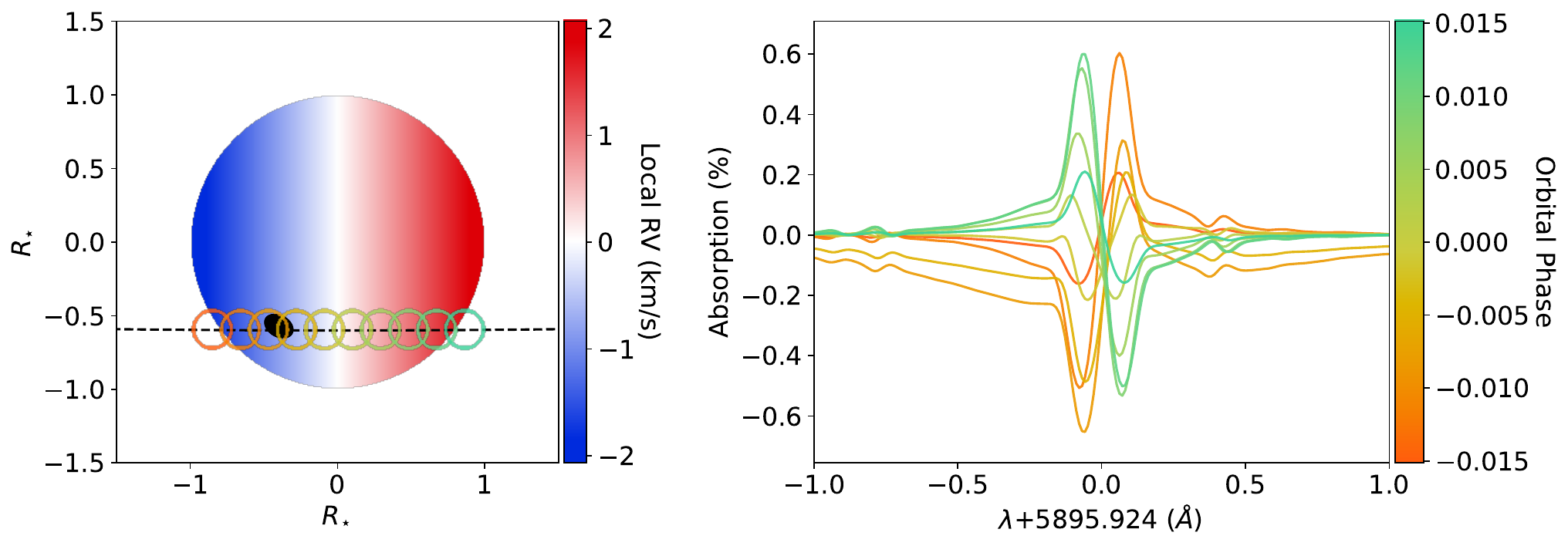}
    \caption{Orbital configuration, spot position, and corresponding absorption spectra for the sodium  D$_1$ line. Left panel: The planetary track is represented by a dashed line, with each open circle corresponding to an absorption spectrum of the same color, highlighting the local distortion. The colormap, ranging from blue to red, represents local velocities. Right panel: Evolution of the absorption spectrum as a function of the orbital phase. The spot-crossing effect induces a slight asymmetry that is visible in the local profiles in the stellar rest frame.}
    \label{fig:hotjupiter}
\end{figure}

\begin{figure}
    \centering
    \includegraphics[width=\linewidth]{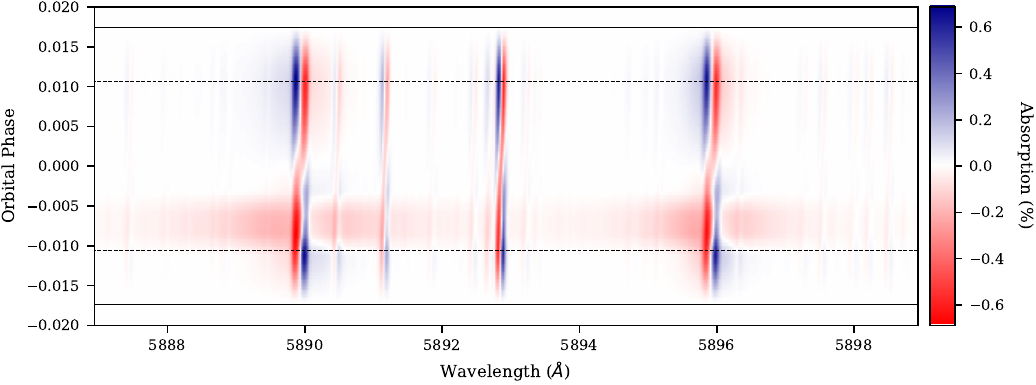}
    \caption{The time-series of absorption spectra around the sodium lines for the spot-crossing scenario is shown. The image covers a wavelength interval of 12~\AA, centered on the mean position between the two lines. Horizontal black lines indicate the transit contacts: solid black represents T1 and T4, while dashed black denotes T2 and T3 (from negative to positive phase values). At phases corresponding to the spot position, an extra distortion is observed due to the spectral difference between the AR and the stellar photosphere. The sodium lines in the AR appear broader than those in the photosphere, resulting in an apparent emission that extends beyond the distortion near the line core.}
    \label{fig:tomography}
\end{figure}
\FloatBarrier
\begin{table}[h!]
\centering
\caption{Solar-analog and Jupiter-analog system properties.}
\label{tab:hotjupiter}
\begin{tabular}{p{5.5cm}c}

\textbf{Properties} & \textbf{Values} \\
\hline
\hline
\\
\multicolumn{2}{c}{\textbf{Stellar}} \\

Radius (R$_{\odot}$) & 1.0 \\
Mass (M$_{\odot}$) & 1.0 \\
$T_{\text{eff}}$ (K) & 5777 \\
Metallicity [Fe/H]& 4.4 \\
$\log g$ (cm s$^{-2}$) & 0.0 \\
P$_\text{rot}$ (days) & 24.47 \\
\\
\multicolumn{2}{c}{\textbf{Planetary}} \\

Radius (R$_*$) & 0.15 \\
Mass (M$_{\oplus}$) & 317.8\\
Semi-major axis (R$_*$) & 9.0  \\
Period (days) & 3.13 \\
Inclination (deg) & 90\tablefootmark{a}\\
$\lambda$ (deg) & 0 \\
b & -0.6\tablefootmark{b}\\
\hline
\end{tabular}
    \tablefoot{We increased the radius ratio relative to the true value, since most hot jupiters are inflated and presented a lower density when compared with Jupiter. For the solar parameters we used as a reference for the effective temperature \citet{1986A&A...159..175N} and for the logarithmic surface gravity \citet{allen1973astrophysical}.\\
        \tablefoottext{a}{Used in Sect. \ref{QS_sun}}
        \tablefoottext{b}{Used in Sect. \ref{AR_sun}}
    }
\end{table}

\begin{figure}[h!]
    \centering
    \includegraphics[width=0.6\linewidth]{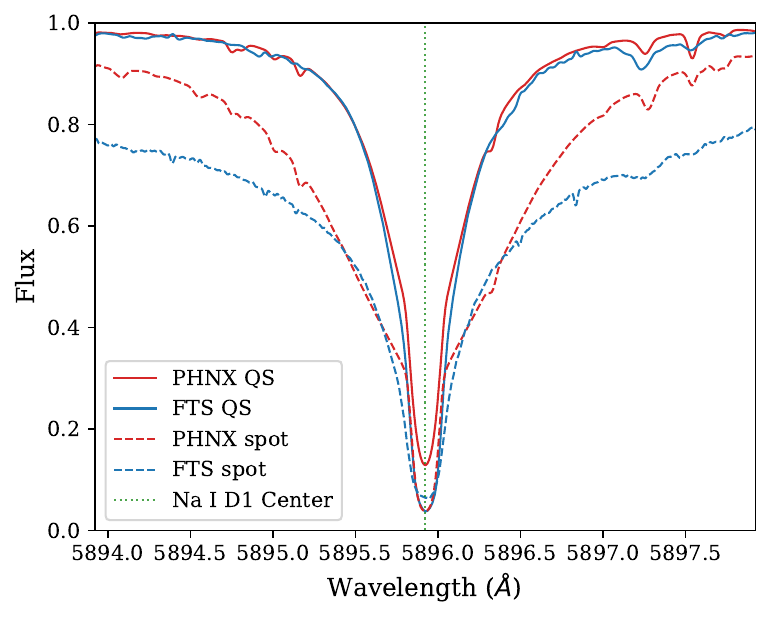}
    \caption{Comparison of the input spectra for the spot-crossing simulation. The figure presents the line profiles of \ion{Na}{I} D$_1$ line from observed spectra obtained with \textsf{FTS-QS} and \textsf{FTS-spot}. These include a quiet Sun observation at disk center \citep{1998assp.book.....W}, shown as solid blue lines, and a sunspot umbral region \citep{2005asus.book.....W}, represented by dashed blue lines. For comparison, we also display \textsf{PHNX} spectra, including the spectrum with the properties of a solar analog for the quiet Sun (solid red) and the spectrum of a star with a temperature 663 K lower than the Sun, representing the active region (dashed red).}
    \label{fig:input_spec_compare}
\end{figure}
\clearpage
\section{HD 209458 Doppler shadow}
\begin{figure}[!htbp]
    \centering
    \includegraphics[scale=1]{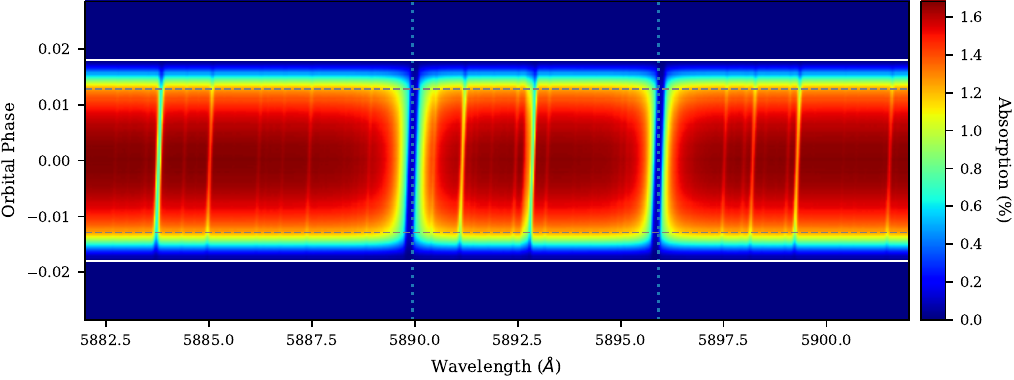}
    \caption{The time evolution of the stellar spectra occulted by the planet as a function of orbital phase, in the wavelength range 5882–5902 $\AA$, which includes the sodium doublet. The positions of the \ion{Na}{I}lines are highlighted by vertical dotted lines. The horizontal white lines indicate the transit contacts, with solid white representing T1 and T4, and dashed gray representing T2 and T3 (from negative to positive values of phase). This map was constructed by subtracting the in-transit flux-weighted spectra from a reference out-of-transit spectrum.}
    \label{fig:shadow}
\end{figure}
\FloatBarrier
\section{Comparison with \texttt{SOAP2.0}}\label{compare_vers}
The version of \texttt{SOAP} implemented in this paper is entirely written in Python, with the most critical and frequently executed parts compiled using Numba \citep{lam2015numba}, a just-in-time compiler for Python. In this section, we compare the activity impact at the CCF level, as described in \citet{2014ApJ...796..132D}, for both a spot and a facula by analyzing the integrated CCF results obtained with the original C code. For these simulations, we set a $1\%$ filling factor region with a temperature contrast of 664 K for the spot and 250.9 K for the facula, following the contrast prescribed by \citet{2010A&A...512A..39M}. The active regions are positioned at the disk center, and the simulation evolves over one rotation period of the star, which has the same properties as those listed in Table \ref{tab:hotjupiter}.\par
\begin{figure}
    \centering
    \includegraphics[width=\linewidth]{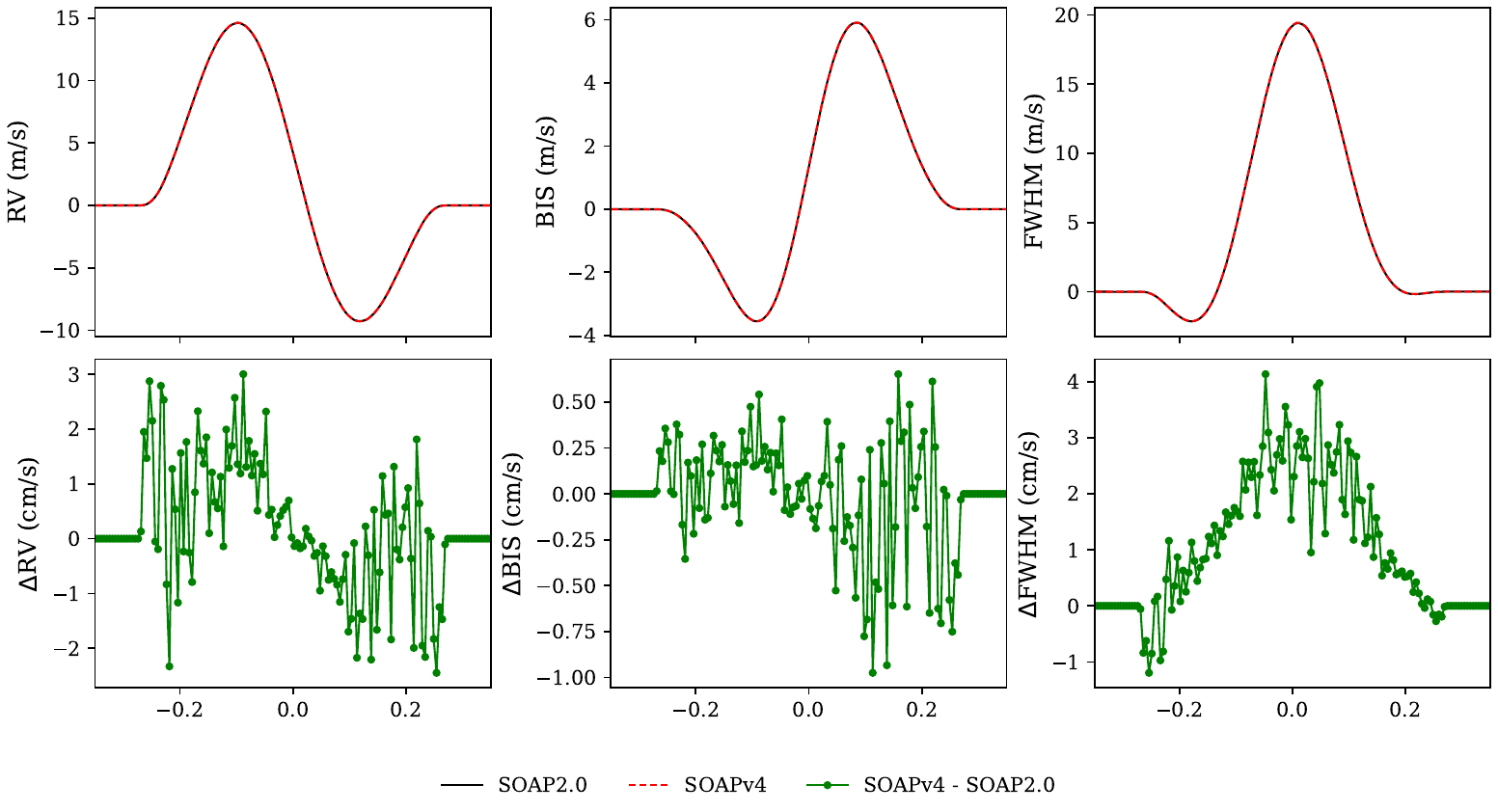}
    \caption{Variation in RV, bisector inverse slope (BIS), and full width at half maximum (FWHM) as a function of stellar rotation phase for an equatorial stellar spot with a filling factor of 1$\%$ and a temperature contrast of 663 K below the photosphere. Top: Comparison of the cross-correlation function (CCF) properties simulated using \texttt{SOAP2.0} and \texttt{SOAPv4}. Bottom: Difference between the simulations obtained with \texttt{SOAPv4} and \texttt{SOAP2.0}.}
    \label{fig:soap2_comparison_spot}
\end{figure}
\begin{figure}
    \centering
    \includegraphics[width=\linewidth]{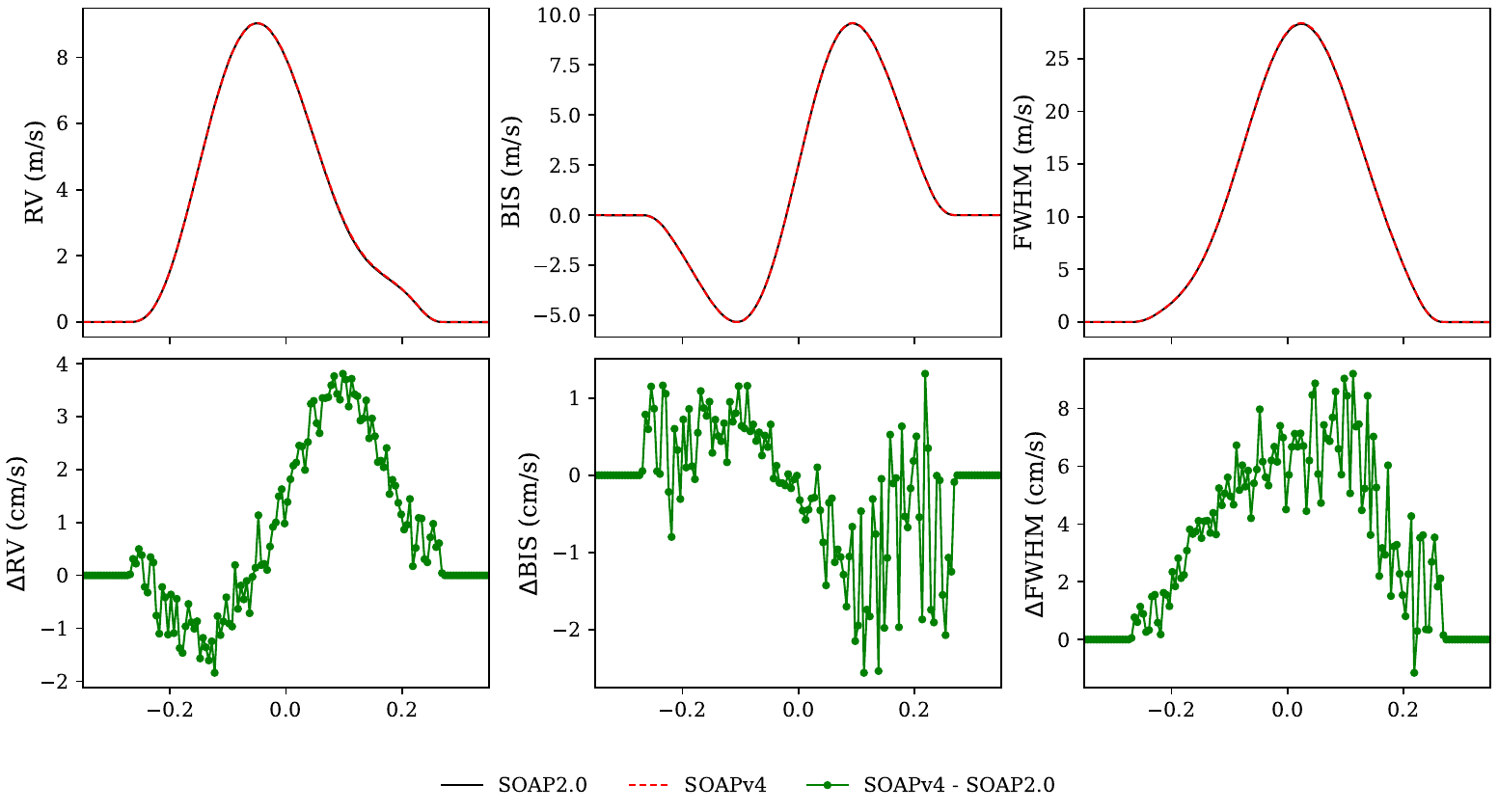}
    \caption{Same CCF properties and descriptions as in Fig. \ref{fig:soap2_comparison_spot}, but for a facula with a stellar disk-center to limb temperature contrast of 250.9 K above the photosphere.}
    \label{fig:soap2_comparison_facula}
\end{figure}

A comparison of the two versions, along with the residuals, is shown in Fig. \ref{fig:soap2_comparison_spot} for a spot and Fig. \ref{fig:soap2_comparison_facula} for a facula. The variation induced by equatorial spots and faculae agrees well, although residuals at the \cms level are present.\par
The differences between the current model and the previous one primarily stem from two factors: an enhanced interpolation algorithm for CCF shifts during the construction of the stellar and AR signals, and a more refined method for selecting pixels in the grid. In the previous model, the interpolation process was carried out in two steps: first, the integer-pixel shift of the CCF was determined, and second, a first-order finite difference approximation was applied to the fractional remainder. This approach has now been replaced with a linear piecewise interpolation method, which offers improved performance, particularly for larger velocity shifts of the fractional remainder. Regarding pixel selection, the previous version relied on the pixel boundaries, whereas the current model uses the centroid of each pixel. This change enhances the stability of the grid, particularly when selecting an even or odd number of grid points. Additionally, the stability of the total number of pixels for each element in the simulation is improved, as it remains more consistent over the evolution of the simulation.
\clearpage

\end{appendix}
\end{document}